\documentclass[3p,times,twocolumn]{elsarticle}
 \biboptions{comma,sort&compress}
 
\usepackage{graphicx}
\usepackage{amsmath}
\usepackage{here}
\usepackage{ecrc}

\volume{00}

\firstpage{1}

\journalname{Nuclear and Particle Physics Proceedings}
\runauth{Victor Feuillard on behalf of the ALICE collaboration}

\jid{nppp}
\jnltitlelogo{Nuclear and Particle Physics Proceedings}

\usepackage{amssymb}

\usepackage[figuresright]{rotating}

\begin{document}

\begin{frontmatter}

\title{Overview of quarkonium production in ALICE} 
 
 \cortext[cor0]{Review talk presented at QCD22, 25h International Conference in QCD (04-07/07/2022, Montpellier - FR). }

 \author[label1]{Victor Feuillard on behalf of the ALICE collaboration}
 \address[label1]{Physikalisches Institut, Im Neuenheimer Feld 226, 69120 Heidelberg, Germany}
\ead{victor.jose.gaston.feuillard@cern.ch}

\pagestyle{myheadings}
\markright{ }
\begin{abstract}
\noindent
ALICE is an experiment mainly dedicated to the study of the quark-gluon plasma (QGP), a state of nuclear matter where quarks and gluons are deconfined, due to high temperature and density. The production of quarkonia, bound states of either a charm and anti-charm quark pair for charmonia or a bottom and anti-bottom quark pair for bottomonia, is a very useful probe of this state of matter. Indeed, the presence of the QGP modifies their production yields, due to a competition between medium-induced suppression and, at least for charmonia, a recombination mechanism in medium or at hadronization.
The measurement of quarkonium production in proton-proton (pp) collisions is also essential to investigate its production mechanism as well as to test quantum chromodynamics (QCD) models. It also provides a reference for the investigation of the properties of the QGP in nucleus--nucleus collisions. Measurements in proton--nucleus collisions allow one to study the cold nuclear matter effects. In addition, high multiplicity pp collisions are useful to investigate multiparton interactions and to search for collectivity in small systems. 
In this contribution, we will report on the latest quarkonium production results, obtained with the ALICE detector in pp collisions, for different center-of-mass energies, at midrapidity and forward rapidity. We will also present the measurements of the quarkonium nuclear modification factor in p--Pb collisions at $\sqrt{s_{NN}}=8.16$~TeV. Finally, in Pb--Pb collisions at $\sqrt{s_{NN}}=5.02$~TeV, we will report on the new $\psi(2S)$ measurement, the nuclear modification factor of the prompt and non-prompt J/$\psi$, as well as the first measurement of quarkonium polarization, the latter carried out also as a function of the event plane. All the measurements will be compared to theoretical predictions.
 
\end{abstract}
\end{frontmatter}
\section{Introduction}

The quark--gluon plasma (QGP) is a state of matter theoretically predicted by quantum chromodynamics (QCD) where quarks and gluons are deconfined. There is a particular interest in studying the QGP since according to current cosmological models our Universe have been in such deconfined state in the early stages after the electroweak phase transition and up to $\tau\approx10$~$\mu$s after the Big Bang. Experimentally, it is possible to create the QGP through ultra-relativistic heavy-ion collisions, such as those happening at the SPS~\cite{Angert:249000}, RHIC~\cite{ROSER200223} or the LHC~\cite{Evans:2008zzb}, but for only a short period of time and within a very small volume (e.g. $\tau\sim$10~fm/\textit{c} and $V\sim5\cdot10\textsuperscript{3}$~fm\textsuperscript{3} in Pb--Pb at $\sqrt s\textsubscript{NN} = 2.76$~TeV)~\cite{Aamodt:2011mr}. Because of their large mass, charm and bottom quarks are primarily produced in hard scatterings at the very beginning of the collision and experience the entire medium evolution. Quarkonium resonances, which are bound states of a $\rm{c\bar{c}}$ pair for charmonium or a $\rm{b\bar{b}}$ pair for bottomonium, are therefore among the most direct signatures for the QGP formation. The properties of the quarkonia can be affected by the QGP in several ways. Firstly, theory predicts that quarkonia are suppressed in a QGP due to the colour screening~\cite{Matsui:1986dk} or dissociation~\cite{PhysRevD.97.014003}. This leads to a reduction of the number of quarkonium states produced in AA collisions with respect to scaled pp collisions. Secondly, a competing mechanism can occur, namely recombination: if there are enough heavy quark pairs produced, then quarkonia can be regenerated by the recombination of these quarks either at the phase boundary~\cite{BraunMunzinger:2000px} or during the QGP phase~\cite{Thews:2000rj}, resulting in an increase of the quarkonium yields. Experimental evidence of the charmonium suppression has been discovered by NA50~\cite{OHNISHI200449}, confirmed by PHENIX~\cite{PHENIX:2011img}, and the interplay between suppression and regeneration phenomena has been put into light by ALICE~\cite{ALICE:2016flj}. Finally, it is argued that the quarkonium polarization can be affected by the strong magnetic field created during the initial stage of the collisions~\cite{Deng:2012pc} as well as by the angular momentum of the medium in non-central collisions~\cite{Liang:2004ph}.

In the following, a selection of the latest quarkonium measurements performed by the ALICE collaboration in pp, p--Pb, and Pb--Pb collisions is reported. Inclusive quarkonia in ALICE are reconstructed down to $p\textsubscript{T} = 0$ at both midrapidity and forward rapidity, through the dielectron and dimuon decay channel respectively. A complete description of the ALICE detector can be found in~\cite{Aamodt:2008zz,Abelev:2014ffa}.

\section{Results in pp collisions}
Quarkonium production measurements in pp collisions provide a reference for the measurement in Pb--Pb collisions. Besides serving as a reference, measurements in pp collisions also allow one to explore phenomenological models, as the quarkonium formation involves both hard-scale processes for heavy quark production, and soft-scale processes for hadronization.

\begin{figure}[htb]
\begin{center}
\vspace{9pt}
\includegraphics[scale=0.24]{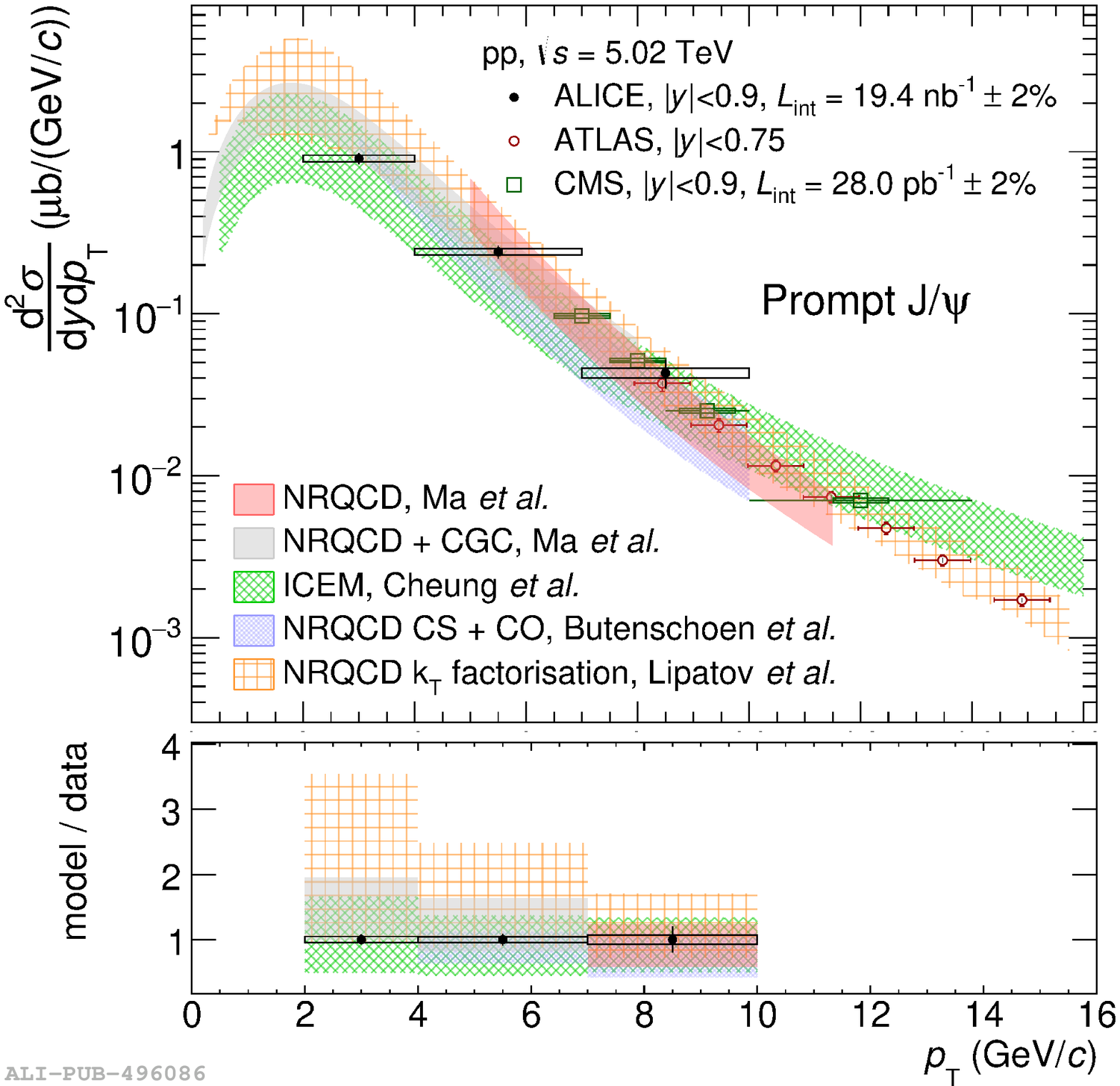}
\includegraphics[scale=0.24]{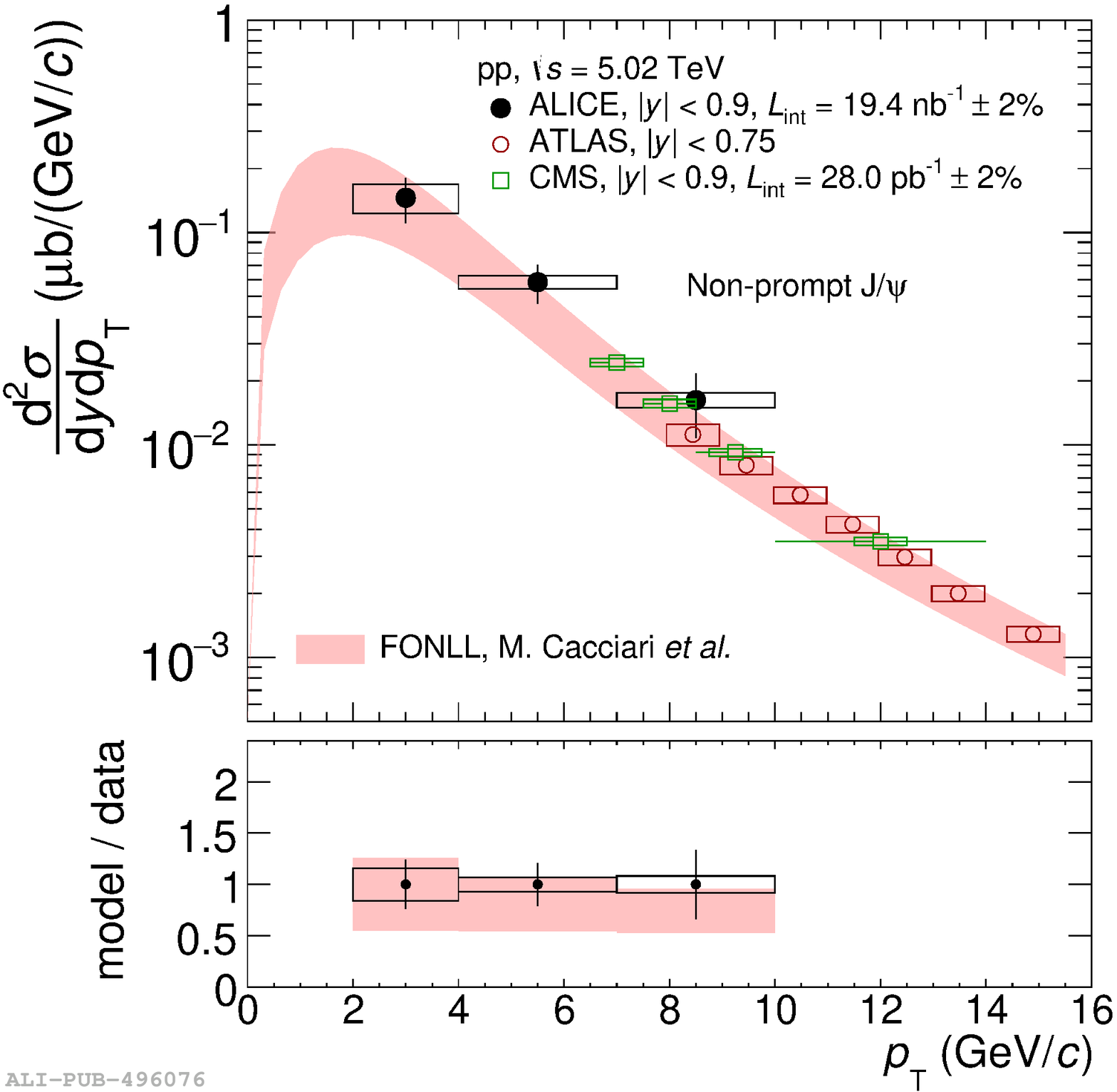}
\caption{J/$\psi$ cross section measured at midrapidity as a function of $p_{\rm{T}}$ for prompt J/$\psi$ (top) and non-prompt J/$\psi$ (bottom)~\cite{ALICE:2021edd}. Results are compared with ATLAS~\cite{ATLAS:2017prf} and CMS~\cite{CMS:2017exb} measurements and model calculations~\cite{PhysRevLett.106.022003,PhysRevLett.106.042002,PhysRevD.100.114021,PhysRevLett.113.192301,Cacciari:2012ny,PhysRevD.98.114029}.}
\label{fig:JPsi_CrossSection}
\end{center}
\end{figure}

Thanks to the excellent spatial resolution provided by the barrel detectors, prompt and non-prompt J/$\psi$ can be disentangled at midrapidity. Recent measurements of prompt and non-prompt J/$\psi$ cross sections in pp collisions at $\sqrt{s} = 5.02$~TeV~\cite{ALICE:2021edd} are presented in Figure~\ref{fig:JPsi_CrossSection}. The cross sections as a function of the transverse momentum ($p\textsubscript{T}$) are shown together with ATLAS~\cite{ATLAS:2017prf} and CMS~\cite{CMS:2017exb} measurements, and compared with several theoretical predictions. In the common $p\textsubscript{T}$ range, the ALICE results agree with the ATLAS and CMS measurements, both for prompt and non-prompt J/$\psi$. 

For prompt J/$\psi$, the evolution of the cross section as a function of $p\textsubscript{T}$ is well described within uncertainties by Non Relativistic QCD~\cite{PhysRevLett.106.022003,PhysRevLett.106.042002,PhysRevD.100.114021,PhysRevLett.113.192301} (NRQCD) and Improved Color Evaporation~\cite{PhysRevD.98.114029} models (ICEM). However, the uncertainties on the model calculations do not allow to discriminate between models. For non-prompt J/$\psi$, the cross section is well described within uncertainties by FONLL calculations~\cite{Cacciari:2012ny}.

\begin{figure}[htb]
\begin{center}
\vspace{9pt}
\includegraphics[scale=0.234]{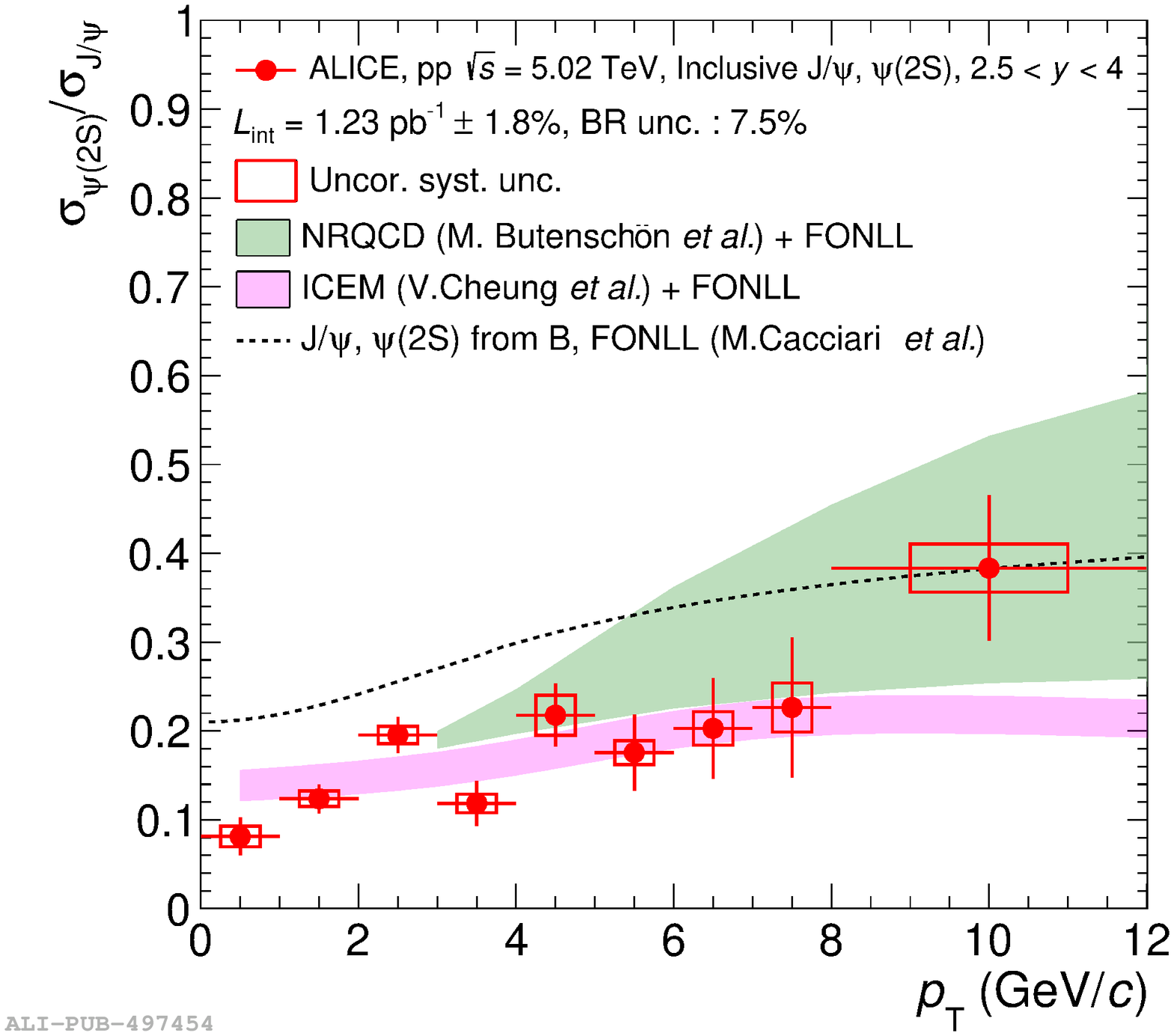}
\includegraphics[scale=0.234]{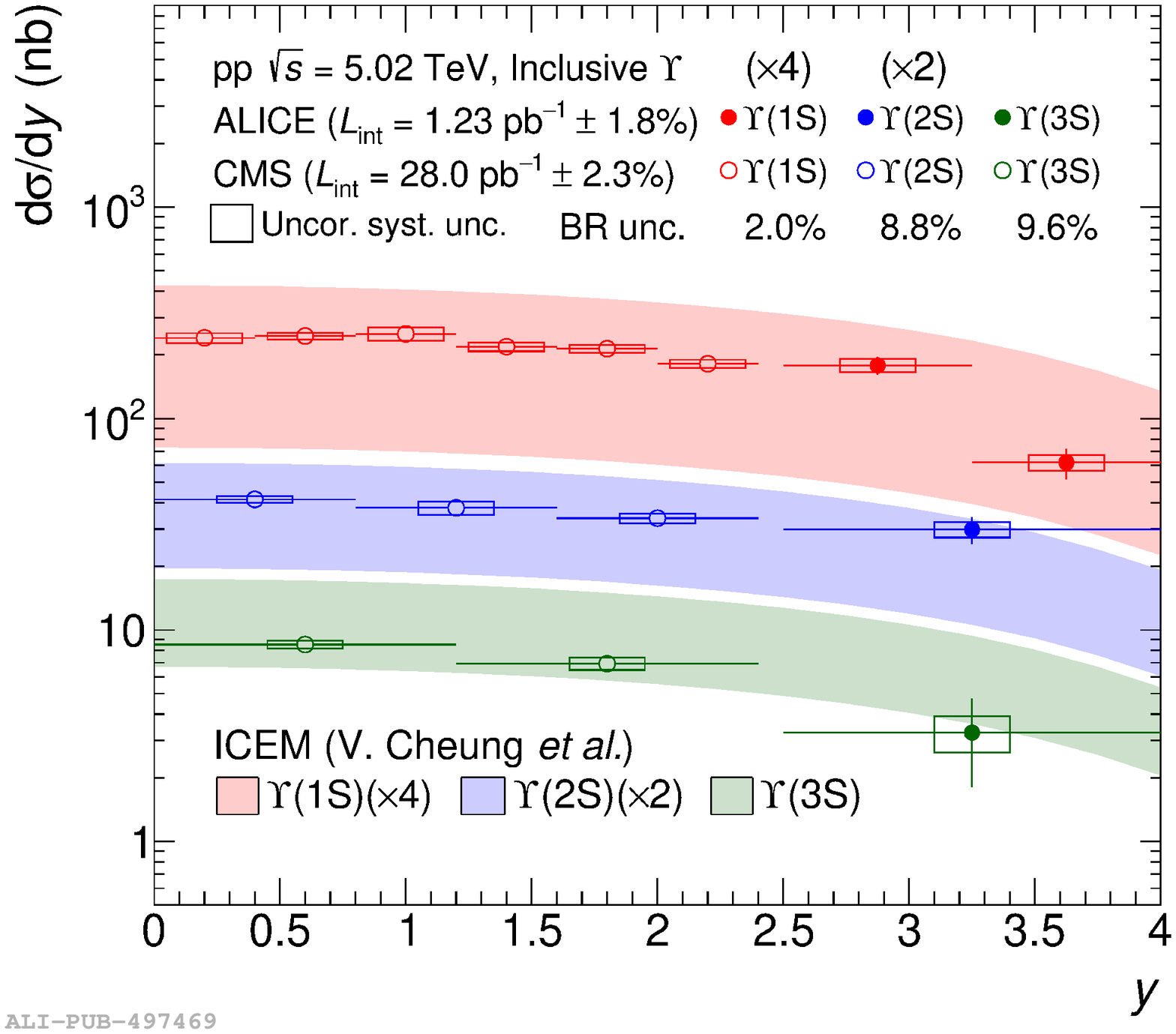}
\caption{Top : $\psi(2S)$-to-J/$\psi$ cross section ratio measured at forward rapidity as a function of $p_{\rm{T}}$~\cite{ALICE:2021qlw} compared with models~\cite{PhysRevLett.106.022003,PhysRevD.98.114029,Cacciari:2012ny}. Bottom : $\Upsilon(nS)$ cross-section as a function of rapidity measured combining ALICE~\cite{ALICE:2021qlw} and CMS~\cite{CMS:2018zza} data and compared with model predictions~\cite{PhysRevD.98.114029}.}
\label{fig:Quarkonia_CrossSection}
\end{center}
\end{figure}

At forward rapidity, only inclusive quarkonia can be measured, and recent measurements in pp collisions at $\sqrt{s} = 5.02$~TeV~\cite{ALICE:2021qlw} are presented in Figure~\ref{fig:Quarkonia_CrossSection}. In particular, the ratio of the inclusive $\psi(2S)$ to J/$\psi$ cross-sections is shown in the top panel where it is compared with NRQCD~\cite{PhysRevLett.106.022003} and ICEM~\cite{PhysRevD.98.114029} model calculations with FONLL~\cite{Cacciari:2012ny} calculations added on top to account for the non-prompt charmonium contribution. Both models describe the ratio within uncertainties. The non-prompt cross section ra- tio from FONLL~\cite{Cacciari:2012ny} is also shown separately for completeness.

The bottom panel of Figure~\ref{fig:Quarkonia_CrossSection} shows the $\Upsilon(nS)$ cross sections as a function of rapidity combining ALICE and CMS data~\cite{CMS:2018zza}, and compared with ICEM predictions~\cite{PhysRevD.98.114029}. For all three resonances, the cross section exhibits a drop-off from a plateau towards forward rapidity, as predicted by the model calculations.

\begin{figure}[htb]
\begin{center}
\vspace{9pt}
\includegraphics[scale=0.315]{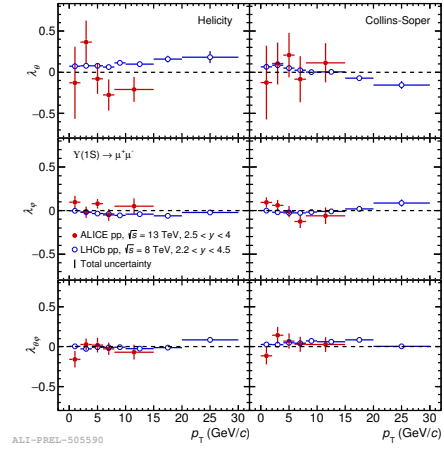}
\caption{$\Upsilon(1S)$ polarization parameters as a function of $p_{\rm{T}}$ measured in the Helicity and the Collins-Soper reference frames. LHCb data are also shown~\cite{LHCb:2017scf}.}
\label{fig:Upsilon_polarization}
\end{center}
\end{figure}

Polarization is defined as the alignment of the particle spin with respect to a chosen axis and can be studied via the polar angle distribution of the dilepton decay products of the quarkonium with respect to a specific quantization axis~\cite{Faccioli:2010kd}. This distribution is parametrized as : $W(\theta) \propto \frac{1}{3+\lambda_\theta} (1+ \lambda_\theta \cos^2 \theta + \lambda_\varphi \sin^2 \theta \cos 2\varphi + \lambda_{\theta\varphi} \sin 2\theta \cos \varphi)$, where $\lambda_\theta$, $\lambda_\varphi$ and $\lambda_{\theta\varphi}$ are the polarization parameters. The first ALICE measurement of the $\Upsilon(1S)$ polarization is shown in Figure~\ref{fig:Upsilon_polarization}, where the polarization parameters are shown as a function of $p\textsubscript{T}$. The analysis is performed considering two different reference systems: the Collins-Soper frame, where the quantization axis is chosen as the bisector between the directions of the two colliding hadrons in the quarkonium rest frame, and the Helicity frame where the quantization axis correspond to the the momentum direction of the $\Upsilon(1S)$ in the center of mass of the colliding system. Within the uncertainties, no polarization is observed neither in the Helicity nor in the Collins-Soper reference frame. These results are compatible with similar measurements from the LHCb collaboration~\cite{LHCb:2017scf}.

\begin{figure}[htb]
\begin{center}
\vspace{9pt}
\includegraphics[scale=0.367]{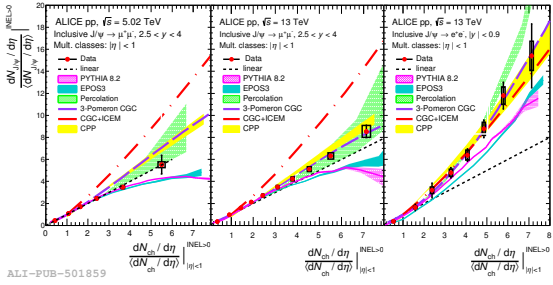}
\caption{Self-normalized J/$\psi$ yield measured at forward rapidity and $\sqrt{s} = 5.02$ TeV(left), at forward rapidity and $\sqrt{s} = 13$ TeV (middle), at midrapidity and $\sqrt{s} = 13$ TeV (right), as a function of the self-normalizes charged particle multiplicity measured at midrapidity, compared with theoretical models~\cite{PhysRevD.101.054023,PhysRevD.98.074025,refId0,PhysRevC.86.034903,Werner:2013tya,SJOSTRAND2015159}.}
\label{fig:Jpsi_vs_mult}
\end{center}
\end{figure}

Finally, the self-normalized J/$\psi$ yields in pp collisions at $\sqrt{s} = 5.02$ and 13 TeV as a function of the self-normalized charged particle multiplicity are shown in Figure~\ref{fig:Jpsi_vs_mult}. Correlations between the J/$\psi$ yield and the charged particle multiplicity are useful to shed light on multiparton interactions, i.e. events where several parton-parton interactions happen in a single hadronic collision. The left and middle panels of Figure~\ref{fig:Jpsi_vs_mult} show the new measurement of the self-normalized J/$\psi$ yields at forward rapidity~\cite{ALICE:2021zkd} at $\sqrt{s} = 5.02$~TeV and 13~TeV, respectively. The right panel displays the self-normalized J/$\psi$ yield at mid-rapidity at $\sqrt{s}=13$~TeV~\cite{ALICE:2020msa}. For the three measurements, the charged particle multplicity is measured at mid-rapidity.

As opposed to midrapidity where the self-normalized J/$\psi$ yield grows faster than linear as a function of the event multiplicity, the J/$\psi$ yield at forward rapidity grows approximatively linearly, independently of the collision energy. These trends can be described qualitatively by models including either initial state effects or both initial and final state effects~\cite{PhysRevD.101.054023,PhysRevD.98.074025,refId0,PhysRevC.86.034903,Werner:2013tya,SJOSTRAND2015159}. Some models, such as the Percolation model, the Coherent Particle Production (CPP) model and the 3 Pomeron Color Glass Condensate (CGC) model even manage to describe the results quantitatively in both rapidity ranges.

\section{Results in p--Pb collisions}
Measurements in p--Pb collisions allow one to study cold nuclear matter (CNM) effects, such as the modifications of the parton distribution functions (PDFs) in nuclei. For a nucleon inside a nucleus, it leads to a change in the probability for a quark or gluon to carry a fraction $x$ of the nucleon momentum. Other effects may come into play, such as energy loss via initial-state radiation~\cite{Bars:2007vv}, energy loss through coherent effects~\cite{Arleo:2014oha}, interactions of the final states with comoving matter~~\cite{FERREIRO201598} or inelastic interactions with the surrounding nucleon~\cite{Kopeliovich:2002yh} (negligible at LHC energies). Results in p-Pb are crucial for the correct interpretation of Pb--Pb measurements, in particular they allow to distinguish the effects originating from the hot medium from those caused by the initial presence of nuclear matter. Furthermore, the asymmetry of the p--Pb collision allows us to probe different Bjorken-$x$ regions of the lead nucleus, in particular measurements in the p-going (Pb-going) direction allow one to investigate low (high) Bjorken-$x$ region.

\begin{figure}[htb]
\begin{center}
\vspace{9pt}
\includegraphics[scale=0.235]{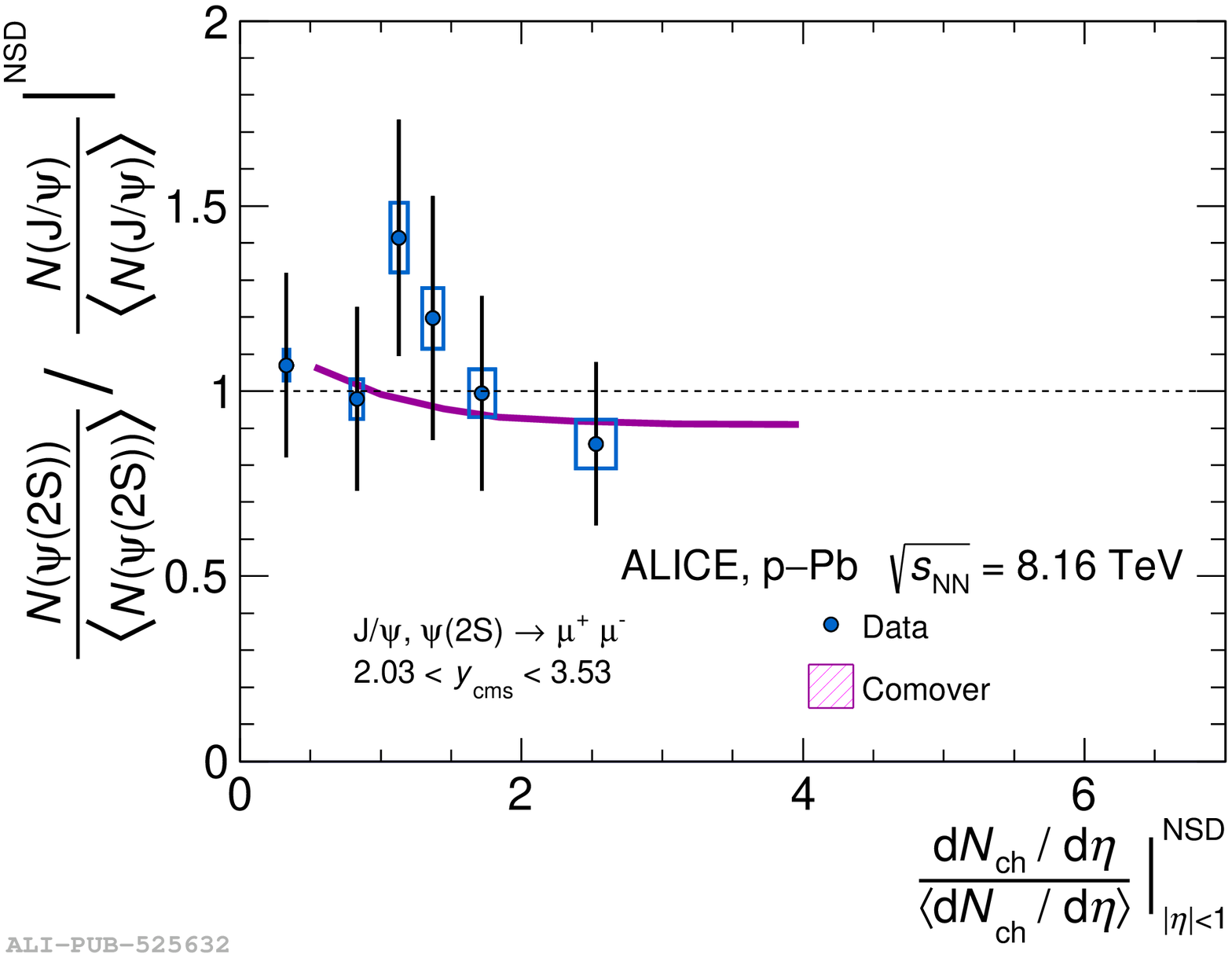}
\includegraphics[scale=0.235]{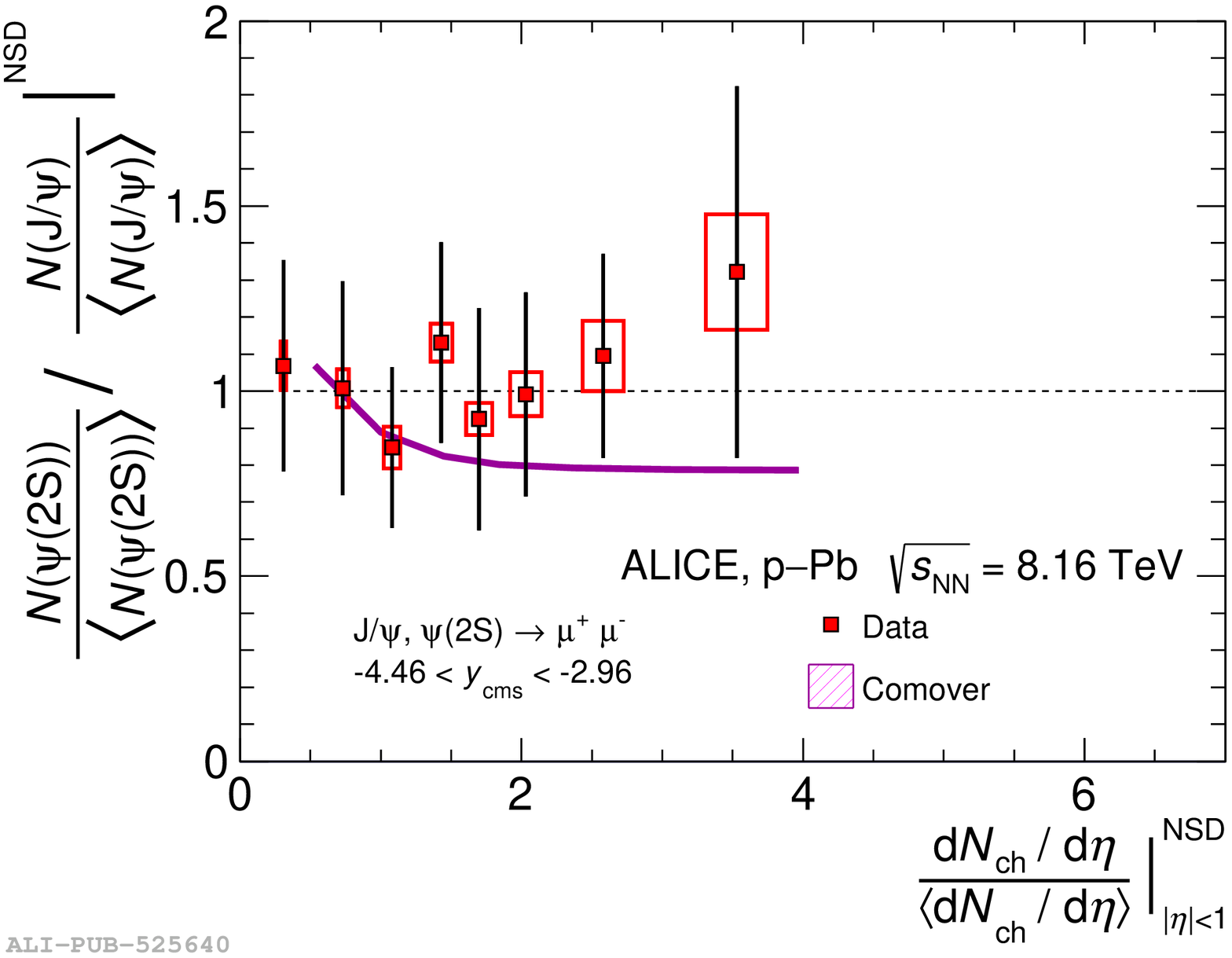}
\caption{Self-normalised $\psi(2S)$-to-J/$\psi$ ratio as a function of midrapidity charged particle multiplicity measured at forward (top) and backward (bottom) rapidity, compared with model predictions~\cite{FERREIRO201598}.}
\label{fig:Psi2S_vs_mult}
\end{center}
\end{figure}

The measurement of correlations between the quarkonium yield and the charged particle multiplicity in p--Pb collisions are useful to understand how initial-state nuclear effects impact the quark-antiquark pair created in the hard scattering, and to investigate final state effects.  
Recent results of the self-normalized $\psi(2S)$ to J/$\psi$ double ratio as a function of the self-normalized charged particle multiplicity in p--Pb collisions at $\sqrt{s_{NN}} = 8.16$~TeV~\cite{ALICE:2022gpu} are shown in Figure~\ref{fig:Psi2S_vs_mult}, at forward (top) and backward (bottom) rapidity. Both in the p-going ($2.03<y_{CMS}<3.53$) and in the Pb-going ($-4.46<y_{CMS}<-2.96$) directions, the double ratio is compatible with unity as a function of the charged particle multiplicity, indicating a similar multiplicity dependence for the $\psi(2S)$ and J/$\psi$ states. This trend can be described by the comover model within the large uncertainties~\cite{FERREIRO201598}, even if the model predicts a ratio $20\%$ below unity at backward rapidity.

\begin{figure}[htb]
\begin{center}
\vspace{9pt}
\includegraphics[scale=0.25]{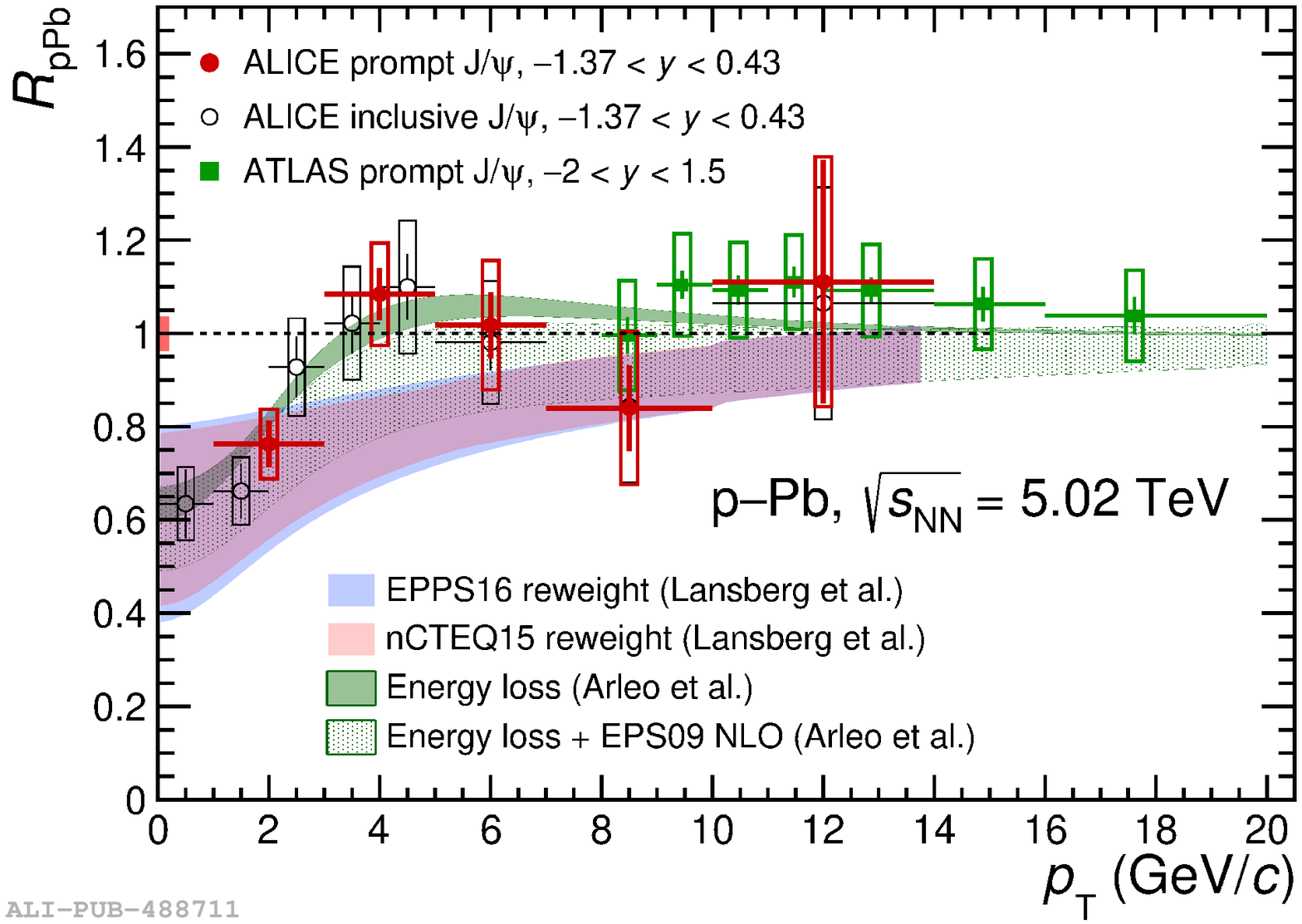}
\includegraphics[scale=0.25]{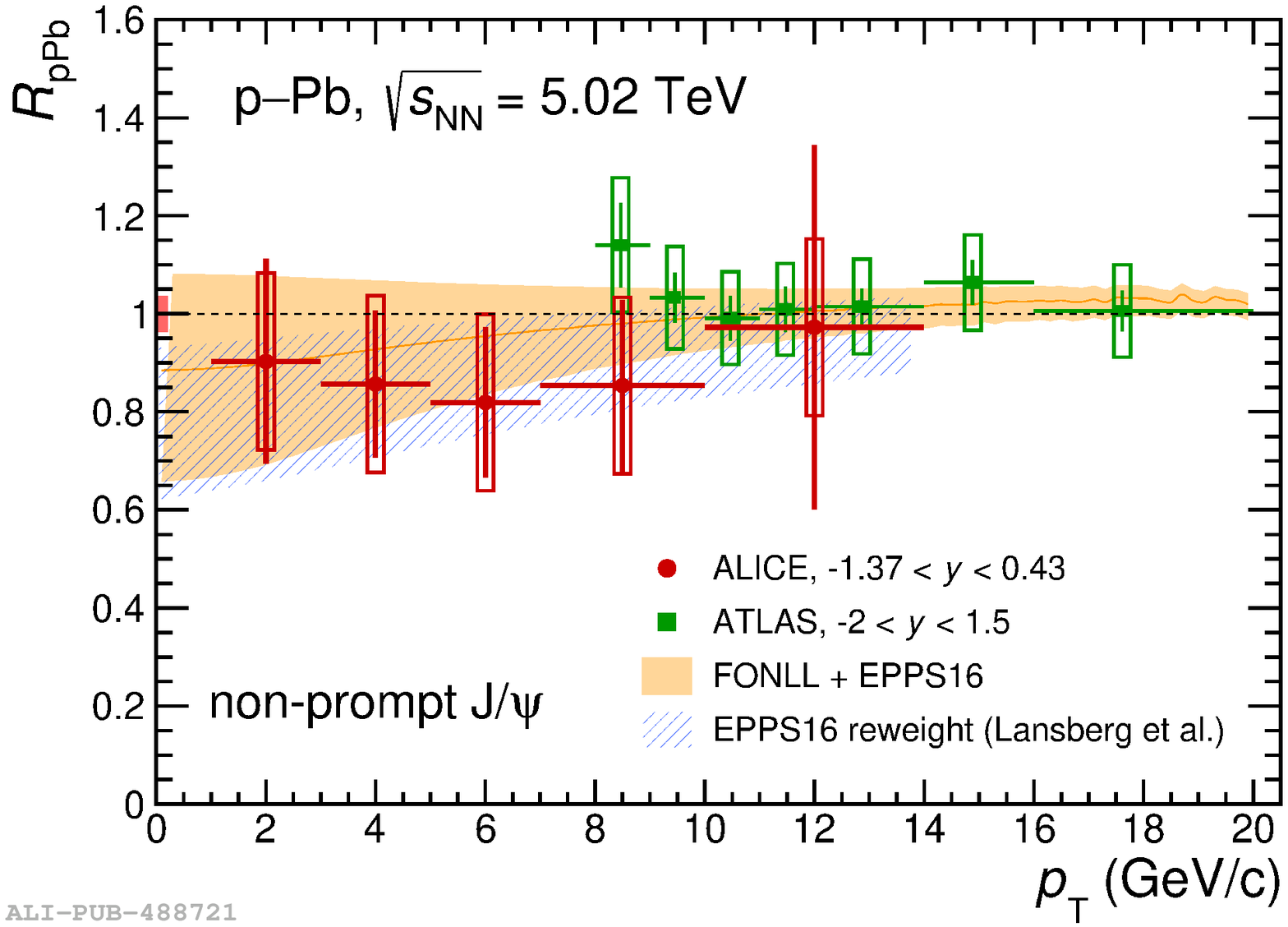}
\caption{$R_{\rm{pPb}}$ measured at midrapidity~\cite{ALICE:2021lmn} as a function of $p_{\rm{T}}$ for inclusive and prompt J/$\psi$ (top) and non-prompt J/$\psi$ (bottom) compared with theoretical predictions~\cite{Arleo:2013zua,PhysRevLett.121.052004,Eskola:2016oht} and ATLAS measurements~\cite{ATLAS:2017prf}.}
\label{fig:JPsi_RpPb}
\end{center}
\end{figure}

CNM effects can also affect the quarkonium production, which can be investigated by measuring the nuclear modification $R\textsubscript{pPb}$ in p--Pb collisions.  $R\textsubscript{pPb}$ is defined as the ratio of the production yield in p--Pb with the production yield in pp collisions normalized to the number of binary collisions. A new measurement of the nuclear modification factor $R\textsubscript{pPb}$ measured in p--Pb collisions at $\sqrt{s_{NN}} = 5.02$~TeV~\cite{ALICE:2021lmn} is shown in Figure~\ref{fig:JPsi_RpPb} for the inclusive, prompt (top panel) and non-prompt J/$\psi$ (bottom panel) in the midrapidity region. For inclusive and prompt J/$\psi$, $R\textsubscript{pPb}$ exhibits a similar trend, with a clear suppression for $p\textsubscript{T} < 3$~GeV/$c$. For non-prompt J/$\psi$, a smaller suppression compared to prompt J/$\psi$ is observed, with no $p\textsubscript{T}$ dependence within uncertainties. In the overlapping $p\textsubscript{T}$ range, ALICE results agree with the ATLAS measurements~\cite{ATLAS:2017prf}. Theoretical calculations~\cite{Arleo:2013zua,PhysRevLett.121.052004,Eskola:2016oht}, incorporating various combinations of CNM effects, reproduce the data within uncertainties.

\section{Results in Pb--Pb collisions}

\begin{figure}[htb]
\begin{center}
\vspace{9pt}
\includegraphics[scale=0.23]{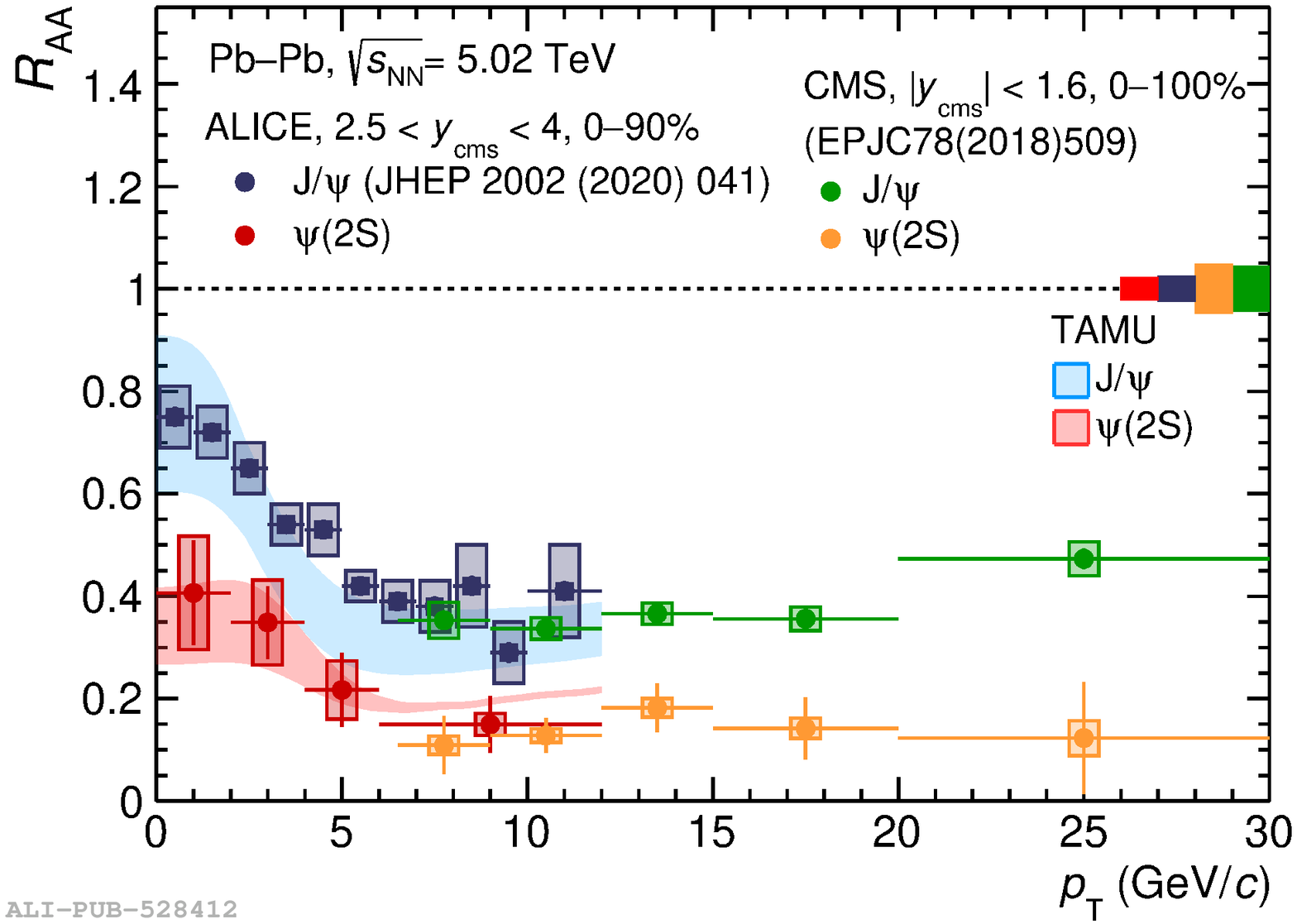}
\includegraphics[scale=0.23]{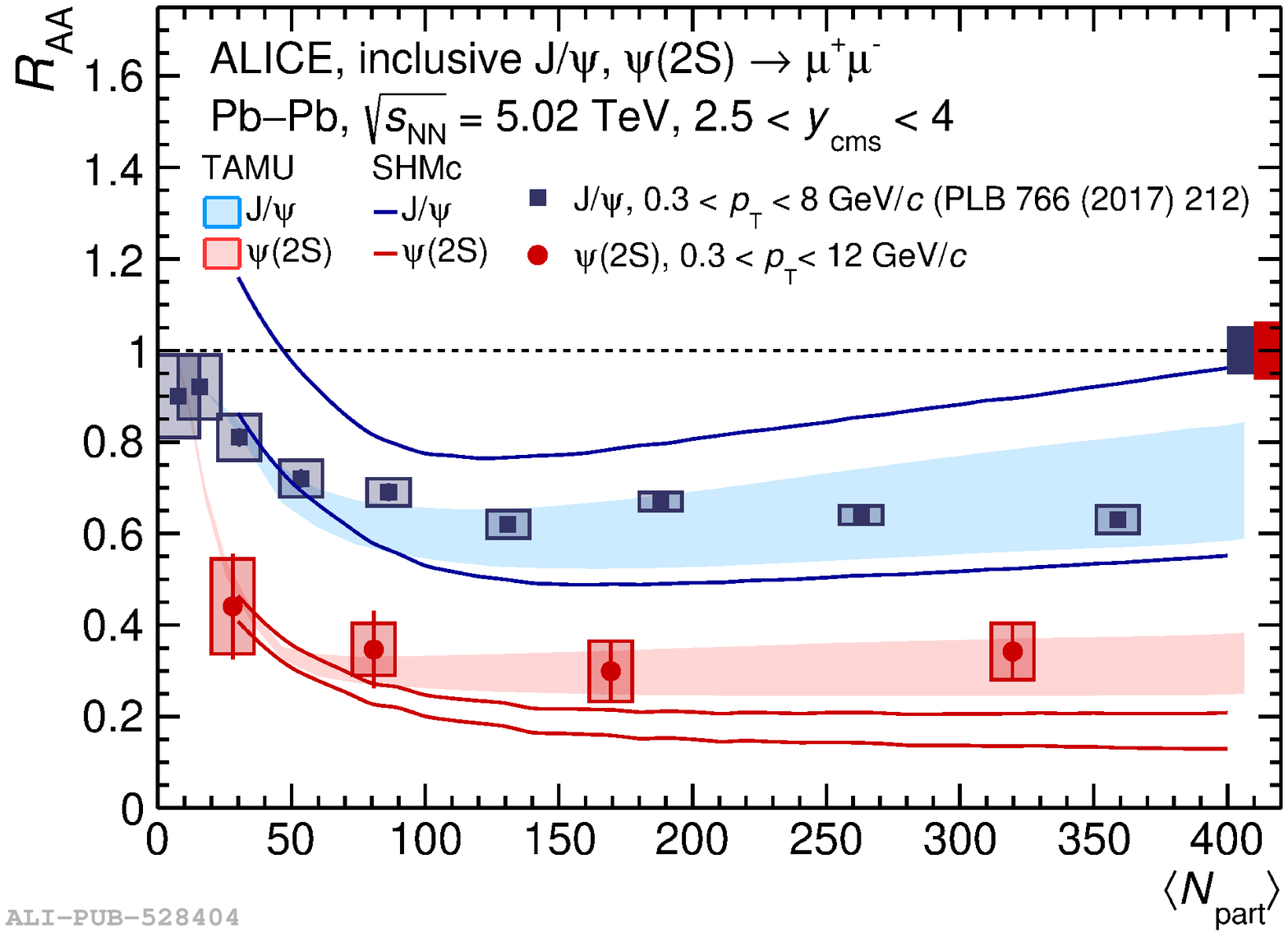}
\caption{Top : J/$\psi$ and $\psi(2S)$ $R_{\rm{AA}}$ measured at forward rapidity as a function of $p_{\rm{T}}$~\cite{ALICE:2022jeh} together with CMS measurements~\cite{CMS:2017uuv}, and compared with theoretical predictions~\cite{DU2015147}. Bottom :  J/$\psi$ and $\psi(2S)$ $R_{\rm{AA}}$ measured at forward rapidity as a function of centrality~\cite{ALICE:2022jeh}. Results are compared with model calculations ~\cite{DU2015147,Andronic:2019wva}.}
\label{fig:Psi2S_RAA}
\end{center}
\end{figure}

The $p\textsubscript{T}$-dependent $\psi(2S)$ nuclear modification factor ($R\textsubscript{AA}$) for Pb-Pb collisions at $\sqrt{s_{NN}} = 5.02$~TeV and at forward rapidity~\cite{ALICE:2022jeh} is shown in the top panel of Figure~\ref{fig:Psi2S_RAA}, and compared to the J/$\psi$ result. A strong increase can be observed for both resonance states towards low-$p\textsubscript{T}$, which is an indication of recombination. The ALICE results are compared with the corresponding results from CMS~\cite{CMS:2017uuv} and show a good agreement in the overlapping $p\textsubscript{T}$ range, despite the different rapidity coverage. The transport model~\cite{DU2015147} predictions are in good agreement with the data as a function of $p\textsubscript{T}$.

As shown in the bottom panel of Figure~\ref{fig:Psi2S_RAA}, the results as a function of centrality indicate that the $\psi(2S)$ is more suppressed than the J/$\psi$ in the entire centrality range, and the $\psi(2S)$ $R\textsubscript{AA}$ does not exhibit any trend within the current uncertainties. The results as a function of centrality are compared with transport model and SHMc predictions~\cite{Andronic:2019wva}. Both models reproduce the data within uncertainties, however, SHMc predictions tend to slightly underestimate the data in central collisions.

\begin{figure}[htb]
\begin{center}
\vspace{9pt}
\includegraphics[scale=0.305]{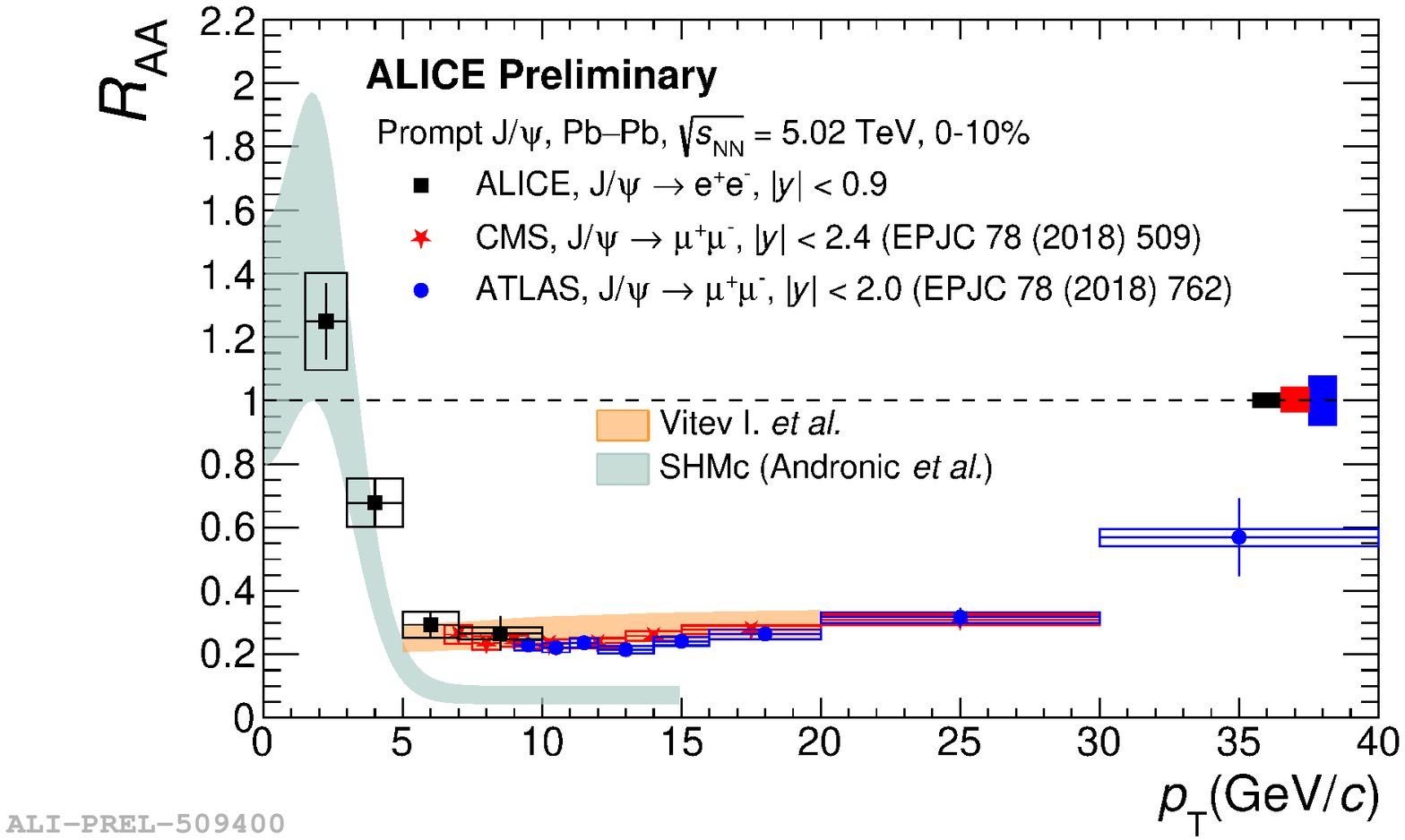}
\includegraphics[scale=0.305]{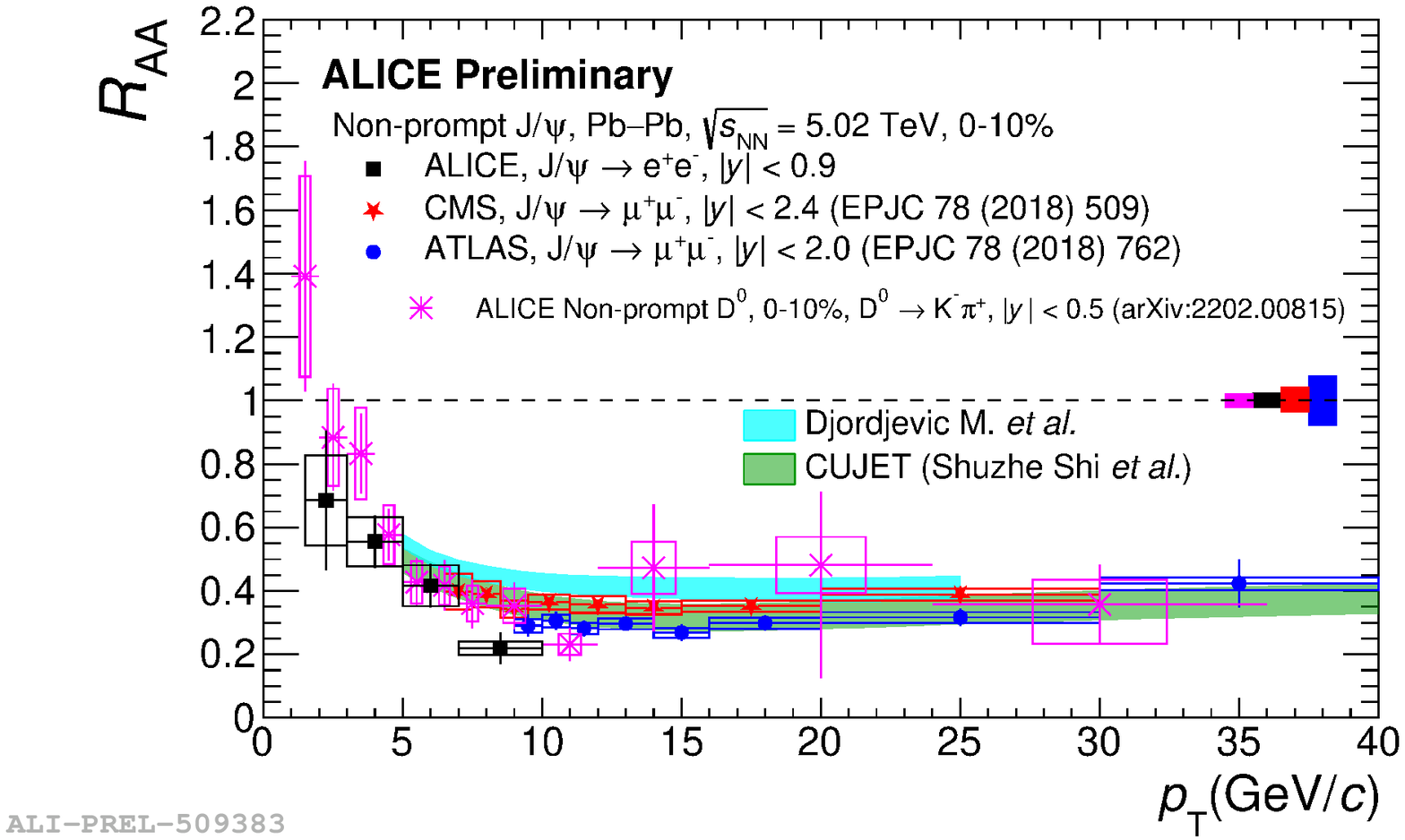}
\caption{Top :  prompt J/$\psi$ $R_{\rm{AA}}$ as a function of $p\textsubscript{T}$, compared with ATLAS~\cite{ATLAS:2018hqe} and CMS~\cite{CMS:2017uuv} measurements and with theoretical models~\cite{Andronic:2021erx,Makris:2019ttx}. Bottom : non-prompt J/$\psi$ $R_{\rm{AA}}$ as a function of $p\textsubscript{T}$, compared with ATLAS~\cite{ATLAS:2018hqe} and CMS~\cite{CMS:2017uuv} measurements and with theoretical models of b-quark energy loss in a QGP~\cite{Shi_2019,Zigic:2021fgf}. Also shown is the $R_{\rm{AA}}$ for ALICE non-prompt $D^0$ mesons.}
\label{fig:JPsi_RAA}
\end{center}
\end{figure}

The J/$\psi$ nuclear modification factors for prompt and non-prompt J/$\psi$ measured in Pb--Pb collisions at $\sqrt{s_{NN}}=5.02$~TeV and at midrapidity are shown in the top and bottom panel of Figure~\ref{fig:JPsi_RAA}, for central collions and as a function of $p\textsubscript{T}$. The prompt J/$\psi$ are not suppressed at low $p\textsubscript{T}$ but only at higher $p\textsubscript{T}$, indicating the dominance of recombination mechanism at low-$p\textsubscript{T}$ in the charmonium sector. As shown in the top panel of Figure~\ref{fig:JPsi_RAA}, the $p\textsubscript{T}$-dependent prompt J/$\psi$ $R\textsubscript{AA}$ is compatible with SHMc model predictions~\cite{Andronic:2021erx} up to 5 GeV$/c$ while at higher $p\textsubscript{T}$, the data is compatible with a model embedding NRQCD into a highly dense gluonic medium~\cite{Makris:2019ttx} whereas SHMc underpredicts the data. For non-prompt J/$\psi$, the data is compatible with models implementing beauty quark energy loss including both collisional and radiative processes~\cite{Shi_2019,Zigic:2021fgf} at high $p\textsubscript{T}$. Compared with ATLAS~\cite{ATLAS:2018hqe} and CMS~\cite{CMS:2017uuv} data, the ALICE data, including non-prompt $D^0$ mesons, are compatible in the overlapping $p\textsubscript{T}$ range.

\begin{figure}[htb]
\begin{center}
\vspace{9pt}
\includegraphics[scale=0.26]{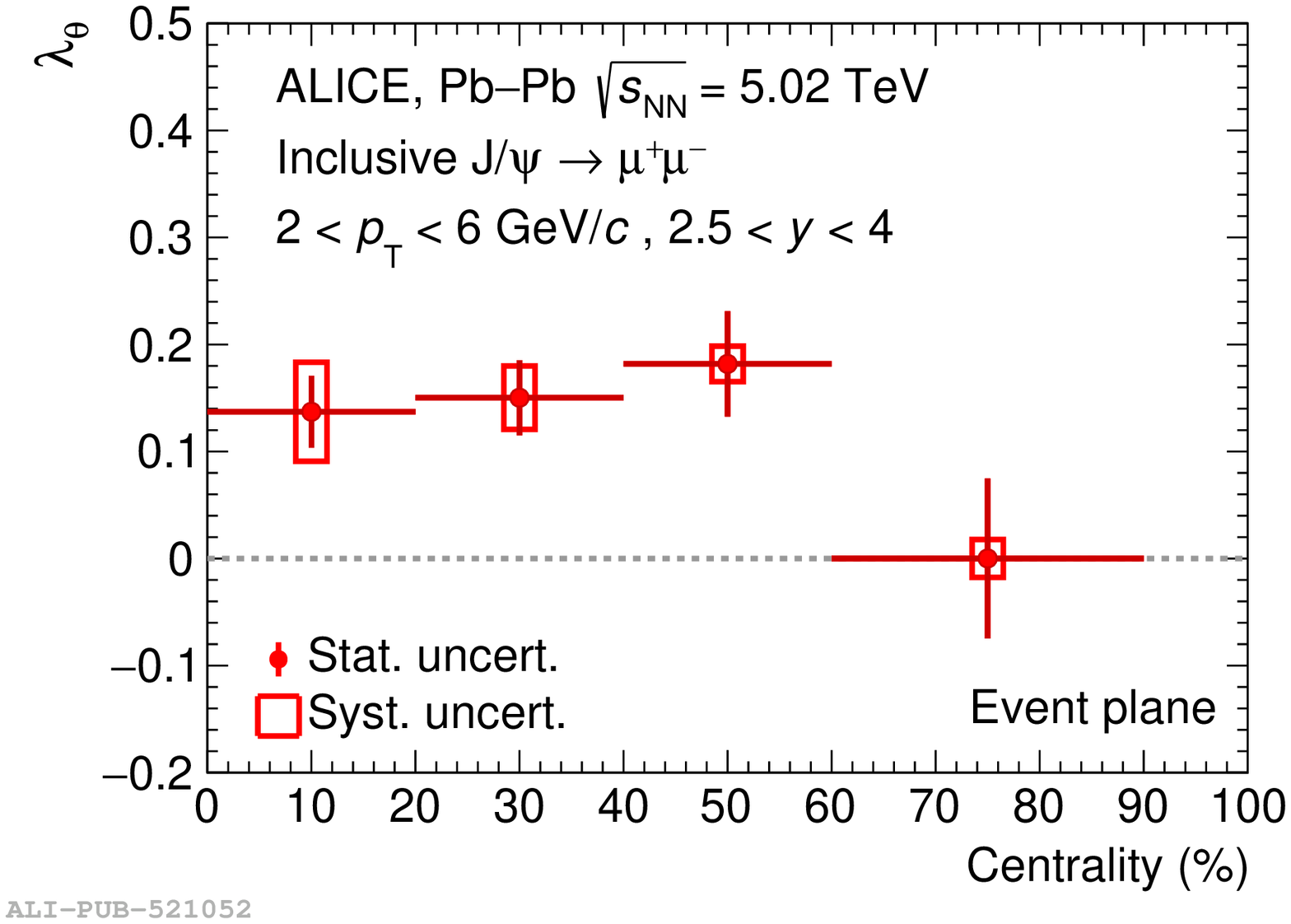}
\includegraphics[scale=0.26]{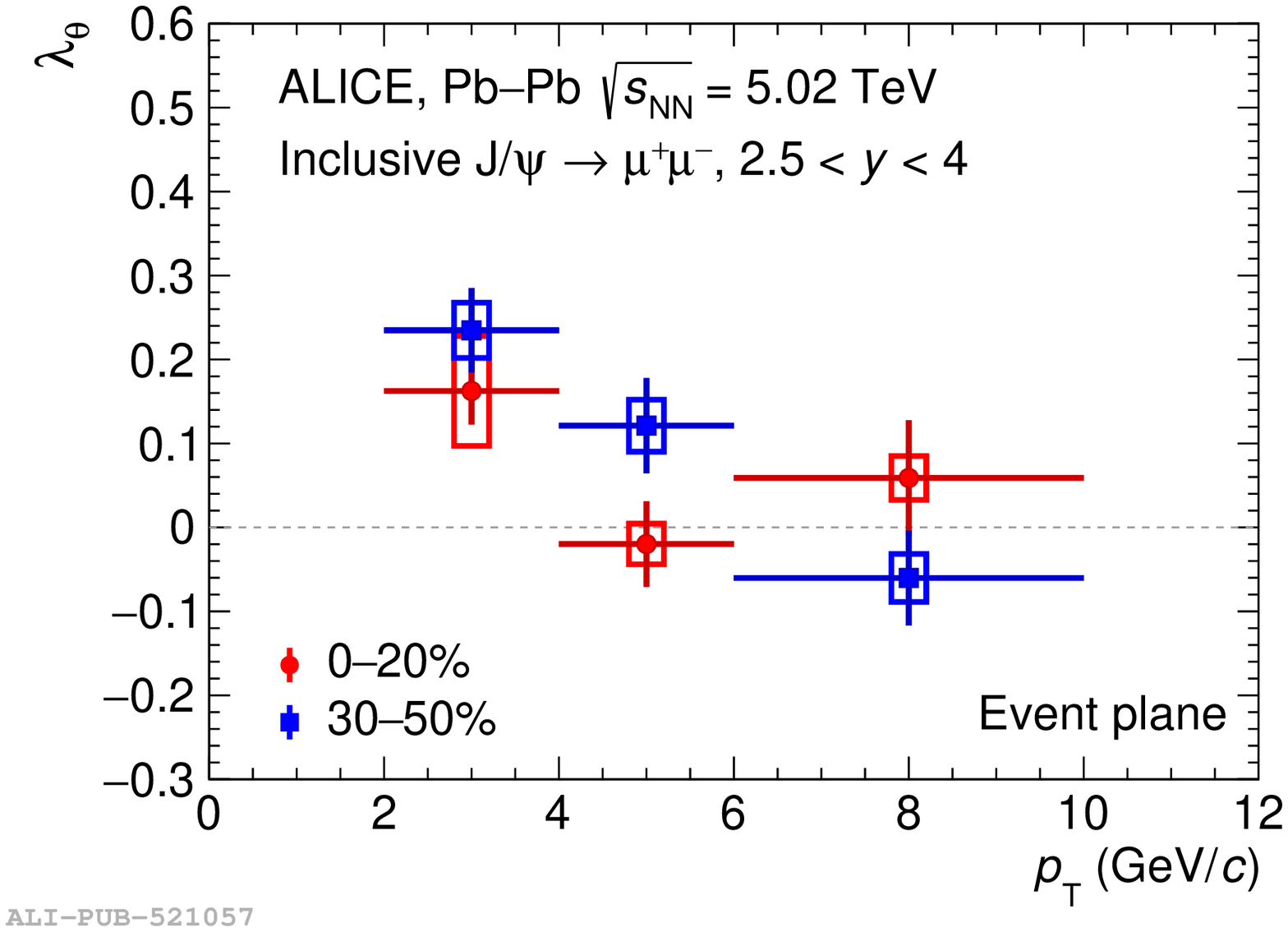}
\caption{J/$\psi$ polarization parameters as a function of centrality (top) and $p_{\rm{T}}$ (bottom) for two centrality ranges, computed relative to the event plane~\cite{ALICE:2022sli}.}
\label{fig:JPsi_polarization}
\end{center}
\end{figure}

Finally, the polarization of the J/$\psi$ has been measured in Pb--Pb collisions as a function of the event plane~\cite{ALICE:2022sli}. The results of $\lambda_\theta$ as a function of centrality and $p\textsubscript{T}$ are shown in Figure~\ref{fig:JPsi_polarization}. A significant non-zero polarization is found in central and semi-central collisions. The results as a function of $p\textsubscript{T}$ indicate that the deviation from zero is larger at small transverse momentum. This tends to show that effects at play in the early stages of the collision, such as magnetic field, contribute less to the polarization, but no theoretical predictions are available to confront these very interesting data at the moment.

\section{Conclusion and Outlook}

An overview of some of the latest results on quarkonium production in ALICE has been presented, in pp, p--Pb and Pb--Pb collisions. These data have been confronted to models when available, providing additional experimental constraints on quarkonium production mechanisms in vacuum and in the QGP. These results will be improved upon and completed by the measurement of new observables during Run 3, thanks to the installation of new and upgraded detectors in ALICE.

\nocite{*}
\bibliographystyle{elsarticle-num}
\bibliography{Feuillard_V.bib}

\end{document}